\documentclass[%
reprint,
twocolumn,
showpacs,preprintnumbers,
amsmath,amssymb,
aps,pre, 
nofootinbib,
longbibliography, 
floatfix,
]{revtex4-2}

\usepackage{units}

\usepackage{upgreek}
\usepackage{textcomp}
\usepackage{enumitem}  
\usepackage{graphicx}
\usepackage{epstopdf}
\usepackage{dcolumn}
\usepackage{bm}
\usepackage{hyperref}

\usepackage{xcolor}
\definecolor{mylinkcolor}{rgb}{0.0,0.0,0.66}
\hypersetup{colorlinks=true,linkcolor=mylinkcolor,urlcolor=mylinkcolor,citecolor=mylinkcolor,filecolor=mylinkcolor,breaklinks=true,linktocpage=true,linktoc=all}

\usepackage[normalem]{ulem}
\usepackage{color,soul} 
\usepackage{lipsum}  
\usepackage[export]{adjustbox}

\newcommand{\etal}{\textit{et al.}}

\newcommand{\sdist}{\kern 0.20em}
\renewcommand{\eqref}[1]{Eq.\sdist(\ref{#1})}

\mathchardef\mhyphen="2D

\usepackage{color}

\emergencystretch=\maxdimen
\begin{document}


\title{Pair filamentation and laser scattering in beam-driven QED cascades}
\author{Kenan Qu}
\email[Corresponding author, ]{kq@princeton.edu}
\affiliation{Department of Astrophysical Sciences, Princeton University,  Princeton, New Jersey 08544, USA \looseness=-1 }  
\author{Alec Griffith}
\author{Nathaniel J. Fisch}
\affiliation{Department of Astrophysical Sciences, Princeton University,  Princeton, New Jersey 08544, USA \looseness=-1 }

\date{\today}

\begin{abstract}
	We report the observation of longitudinal filamentation of an electron-positron pair plasma in a beam-driven QED cascade. The filaments are created in the ``pair-reflection'' regime, where the generated pairs are partially stopped and reflected in the strong laser field. The density filaments form near the center of the laser pulse and have diameters similar to the laser wavelength. They develop and saturate within a few laser cycles and do not induce sizable magnetostatic fields. We rule out the onset of two-stream instability or Weibel instability and attribute the origin of pair filamentation to laser ponderomotive forces. The small plasma filaments induce strong scattering of laser energy to large angles, serving as a signature of collective QED plasma dynamics.
\end{abstract}


\maketitle

\section{Introduction} 
The global race to build ultrastrong lasers worldwide~\cite{danson_2019, cartlidge_light_2018,  ELI_2022} has brought us closer to laboratory testing of strong-field QED effects~\cite{ritus_1985, erber_high-energy_1966, di_piazza_extremely_2012}. Although directly producing the Schwinger field, $E_\mathrm{cr}\sim\unit[10^{18}]{Vm^{-1}}$, remains obscure, different schemes~\cite{bell_possibility_2008, sokolov_pair_2010, jirka_electron_2016, Sarri2015, zhu_dense_2016, grismayer_seeded_2017, Lobet_2017, del_sorbo_channel_2019} to magnify the impact of existing technologies have been proposed. The most promising techniques~\cite{bell_possibility_2008, sokolov_pair_2010} involve the collision of an ultrastrong laser pulse with a highly energetic electron beam, boosting the field by a large Lorentz factor in the rest frame of the electron beam to reach the Schwinger limit. 
This method was adopted in the seminal QED experiment at SLAC~\cite{burke_positron_1997,bamber_studies_1999} to yield measurable electron-positron pairs. Similar experiments have been enabled by the development of the laser-wakefield accelerator, which recently reported exploring the quantum radiation reaction using petawatt (PW) lasers and GeV electron beams generated by the laser~\cite{2018_PRXPoder,2018_PRXCole}. 
If the charge in the electron beam can be increased to the nC level using, e.g., a conventional electron accelerator, its collision with multi-PW laser is predicted to create a QED plasma~\cite{Chen2023, zhang_relativistic_2020, meuren2021mp3, meuren2020seminal, Qu_QED2021, Qu_QED2022, Bulanov2005, Griffith2022, Qu_2023PPCF, griffith2023radiation}. 

QED plasma is a state that describes the interplay of both strong-field QED effect and collective plasma dynamics. It plays a key role in extreme astrophysical environments and in experiments using next generation Schwinger-level high power lasers. Thus, validating the QED theory is crucial notwithstanding the significant  challenges associated with generating sufficiently dense electron-positron pairs. Multiple criteria exist to define the threshold where dense pairs transition into a pair plasma. It's noteworthy that  collective effects can manifest in dilute plasmas even if its dimension is shorter than the Debye length and plasma skin depth, as highlighted by Stenson \etal~\cite{stenson_2017}. 
In the context of laser-plasma interaction, plasma effects become important when the plasma frequency constitutes a substantial percent of laser frequency. Indeed, QED PIC simulations~\cite{Qu_QED2021, Qu_QED2022} demonstrate  that creation of pair plasma in the laser field leads to an observable upshift of laser frequency and the amount of frequency upshift is dependent on the plasma frequency~\cite{joshi1990demonstration, wilks1988frequency, mendoncca2000book, Kenan_2018_upshift}. The laser frequency upshift arises from the change in the plasma dispersion relation, which is generally considered as governing the electromagnetic plasma mode.


In this paper, we focus on the spatial properties of the simulation and report  observations of pair plasma filamentation and large-angle laser scattering. 
Pair density filamentation emerges in the ``pair reflection'' regime. This regime is achieved when the laser intensity reaches the threshold value, $I_\mathrm{th}= \unit[10^{22} \mathrm{-} 10^{23}]{W cm^{-2}}$, which is sufficient to slow down and stop the created pair particles through the combined effects of radiation reaction and laser ponderomotive force. When the created pairs are temporarily stopped, they exhibit the lowest Lorentz factor and thus the maximum plasma frequency. The highly inhomogeneous plasma, characterized by a scale comparable the laser wavelength, induces Mie scattering. The scattered light undergoes intensity modulation at different scattering angles, leading to an expulsion of pairs toward low-intensity regions. Simultaneously, the laser is refracted toward regions with low plasma density. These coupled processes induce the ponderomotive filamentation instability~\cite{Kruer1985, Kaw_1973, Young_PRL1988, Sobacchi_2023}. This instability, recognized as one of the fastest growing modes~\cite{BertPRE2004, BretPRL2005} in relativistic plasma streams, is amplified by the reduction of pair slippage rate with the laser pulse, facilitated by pair reflection. Because the pair filaments are automatically aligned with the laser peak, they continuously scatter the laser toward higher angles, causing a quick decrease in peak laser intensity. 
The coherent scattered light, stemming from laser-plasma interaction,  serves as another signature of QED plasma effects. 
The flexibility of detecting the scattering  at large off-axis angles reduces the experimental complexity.

This paper explains the pair filamentation instability through the analysis of results from a 3D QED PIC simulation. The details of the simulation parameters are presented in Sec.~\ref{sec:formation}. The section also describes the evolution of pair filaments, including the pair density, distribution function, and the corresponding magnetic fields. The dynamics of laser pulse is presented in Sec.~\ref{sec:scattering}, in which we provide an analytical estimation of the laser intensity decrease. Finally, we present in Sec.~\ref{sec:conc} our conclusions and discuss the implication of QED plasma experiments.

\begin{figure}[th]
	\centering
	\includegraphics[width=0.8\linewidth,valign=t]{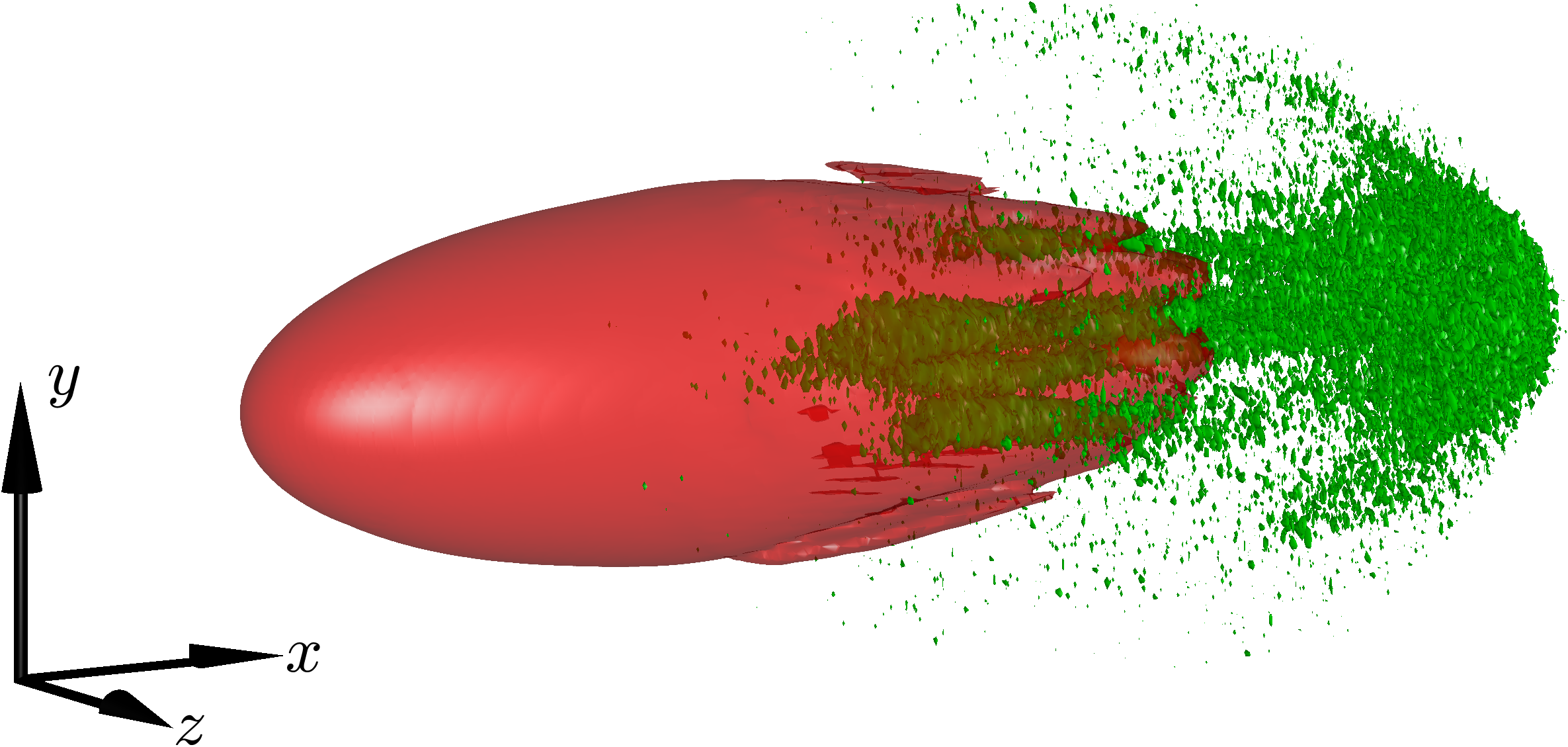}
	\caption{Pair filamentation (green) and laser scattering (red) as a result of $e^-$-beam driven QED cascade. The laser is polarized in the $y$ direction and propagates to $+x$ direction. The $e^-$ beam (not shown) propagates to the $-x$ direction. The snapshot is taken at $t=\unit[0.22]{ps}$. }
	\label{fig:sch}
\end{figure}

\section{Formation of pair filamentation } \label{sec:formation}

The pair filaments are observed in 3D QED PIC simulations that we described in Refs.~\cite{Qu_QED2021,Qu_QED2022}. A snapshot of the laser beam and created pairs are plotted in Fig.~\ref{fig:sch} to show the simulation schematics. The simulations consider  collision of a $\unit[300]{GeV}$ electron beam and a $\unit[24]{PW}$ laser pulse. The electron beam has $\unit[1]{nC}$ charge distributed in a Gaussian sphere with rms radius of $\unit[1]{\mu m}$. Its peak density is $\unit[4\times10^{20}]{cm^{-3}}$. The Gaussian laser pulse has $\lambda= \unit[0.8]{\mu m}$ wavelength,  $\unit[6\times10^{22}]{W cm^{-2}}$ peak intensity (correspondingly $a_0\approx170$), $\unit[50]{fs}$ rms duration, and $\unit[5]{\mu m}$ waist. The laser is linearly polarized in the $y$ direction and propagates to the $-x$ direction. The electron beam propagates to the $+x$ direction. The numerical parameters are similar to those in Ref.~\cite{Qu_QED2021}, but we increase the transverse simulation window size to $(\unit[40]{\mu m})^2$ to capture the pair expansion. The transverse grid size is correspondingly increased to $(\lambda/6)^2$.

The electrons have a maximum quantum parameter $\tilde{\chi}_e \approx220$ at the Gaussian waist in the focal plane, and $\tilde{\chi}_e \approx600$ at the laser focus. 
The high quantum parameter  $\tilde{\chi}_e$ enables a beam-driven QED cascade which creates electron positron pairs through the Breit-Wheeler process. 
The stochastic nature leads to a broad distribution function of the pair momentum with higher creation probability for lower pair energy. Importantly, the low energy pairs play the dominant role in collective plasma dynamics because 
A more detailed description of the QED cascade can be found in Ref.~\cite{Qu_QED2022}. As the pairs continue to lose energy through radiation reaction, those in the low energy spectrum begin to become reflected by the ponderomotive pressure assuming that the laser intensity exceeds the ``pair reflection'' threshold~\cite{Qu_QED2021,Griffith2022}.  Thus, we can divide the pair evolution into three stages, including forward moving, being stopped, and being reflected. All the pairs at different stages can coexist but the pairs in each stage contain both electrons and positrons moving in the same longitudinal direction, making the plasma quasineutral and anisotropic.

\begin{figure}[t]
	\centering
	\includegraphics[width=\linewidth,valign=t]{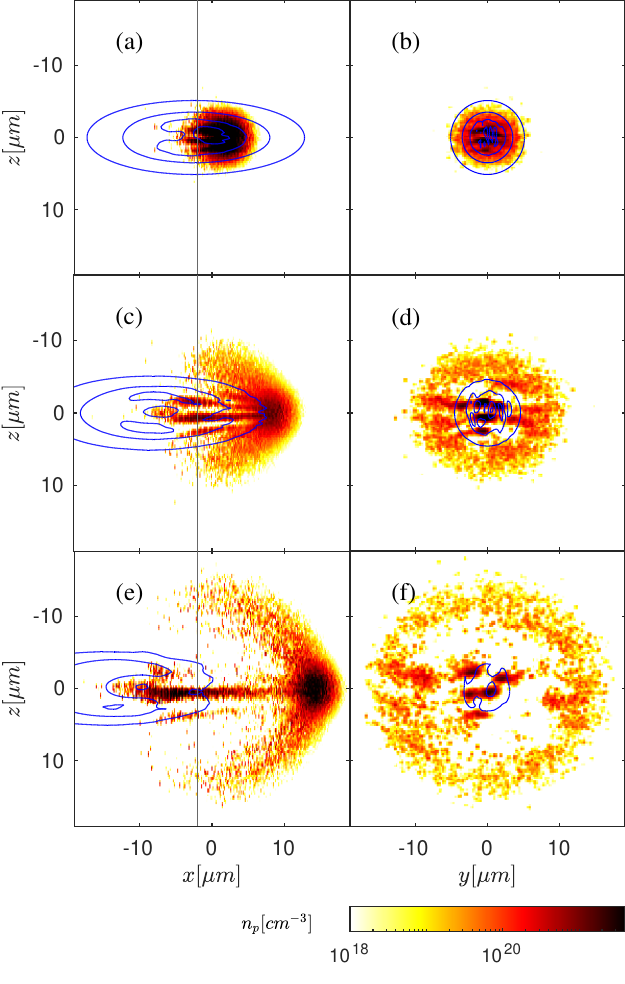}
	\caption{Pair density $(\unit{cm^{-3}})$ at $t=\unit[0.18]{ps}$ (a), (b), $\unit[0.2]{ps}$ (c), (d), and $\unit[0.22]{ps}$ (e), (f), respectively. The blue  curves show the laser intensity contours at $\unit[1,3,5,7\times 10^{22}]{W cm^{-2}}$ from outer to inner, respectively. The left column shows the $y=0$ cross section and the right column shows the $x=\unit[-2]{\mu m}$ cross section indicated by the vertical line on the left. } 
	\label{fig:den}
\end{figure}

The pair plasma, illustrated in green in Fig.~\ref{fig:sch}, exhibits filamentation in the region where it overlaps with the laser pulse. To show more details of the filaments, we plot in Fig.~\ref{fig:den} the evolution of pair density profile and the laser intensity contours in the $x$-$z$ plane at $y=0$ (left column) and the $y$-$z$ plane at $x=\unit[-2]{\mu m}$ (right column). Although the simulation is conducted in a $\unit[100]{\mu m}$-long box, we only show the center part where pair density is finite. Pairs are created with a total charge of $\unit[139]{nC}$ and peak density of $n_p \approx \unit[3\times10^{22}]{W cm^{-2}}$. They initially have a spherical profile similar to the injected electron beam. 
The sphere then expands under the laser radiation pressure and develops density fringes as shown in Fig.~\ref{fig:den}(b). 

With decreasing pair energy, the plasma frequency $\omega_p$ continues to grow. Its peak value is reached when the majority of pairs reach the ``pair reflection'' condition, which is shown in Fig.~\ref{fig:den}(a) as cavitation of the pair sphere at ${t=\unit[0.2]{ps}}$.  The laser field also shows distortion when propagating through the dense plasma. 
The maximum plasma frequency corresponds to a skin depth $c/\omega_p \approx \unit[12]{\mu m}$. 
The fringes condense into filaments and gain a wave vector in the $y$ direction, which can be seen in Fig.~\ref{fig:den}(d). As the same time, the cavitation continues to expand. The pair plasma develops three structures, including a dense core in the front, an expanding shell following the core, and five observable filaments inside the shell. All the structures remain through the rest of the interaction. Interestingly, filamentation allows the pairs to maintain a high center density despite fast expansion of the shell. 


Filamentation of plasmas could arise from or be influenced by two-stream instability, Weibel instability~\cite{BretPRL2005, BertPRE2004, Angelo_2015}, or ponderomotive filamentation instability~\cite{Kruer1985, Epperlein1990}. 
Although the formation of pair filamentation is correlated to counterpropagating pair streams, we cannot directly attribute the filamentation to the plasma streaming instability. 
The streaming instability exists in plasmas when counterpropagating streams carry a longitudinal current and a ``return current'' which produces strong transverse magnetic field to focus the streams~\cite{Weibel_1959,Harris_1959,Fried_1959}. 
However, the electron positron pairs respond symmetrically to the longitudinal laser pressure, including both the radiation reaction and the laser ponderomotive force. The longitudinally flowing pair streams are charge neutral and current neutral, at least in the center, and cannot produce a current or a ``return current.'' Thus, the laser ponderomotive force is not likely to produce streaming instabilities of pair plasmas. In the specific simulation, the filaments grow and saturate within ten laser cycles. It is much shorter than the time scales~\cite{BretPRL2005, BertPRE2004} of two-stream instability $\tau_\mathrm{TS} \sim \gamma/\omega_p \sim 2300/\omega_0$ or the Weibel instability $\tau_\mathrm{W} \sim \sqrt\gamma/\omega_p \sim 95/\omega_0$, where $\omega_0$ is the laser frequency and $\gamma$ is the pair Lorentz factor. 

\begin{figure}[thb]
	\centering
	\includegraphics[width=\linewidth,valign=t]{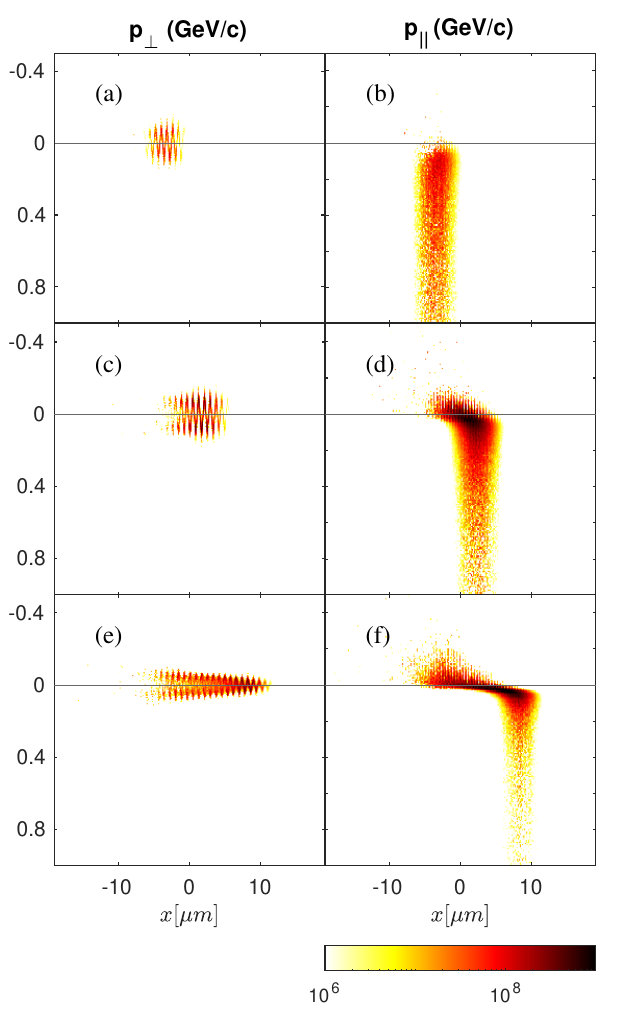}
	\caption{Pair momentum distributions in the transverse direction (left column) and longitudinal direction (right column) at $t=\unit[0.16]{ps}$ (a), (b), $\unit[0.18]{ps}$ (c), (d), and $\unit[0.2]{ps}$ (e), (f), respectively.} 
	\label{fig:distr}
\end{figure}

More details of the filamentation process are shown in the pair momentum space in Fig.~\ref{fig:distr}. The transverse momentum distribution $p_y$, plotted in the left column, reveals the pair oscillation in the laser field. Interestingly, the distribution in $p_y$ shows narrow width compared to the maximum momentum. This can be explained from the factor that the creation and deceleration of pairs are synchronized with the peak laser amplitude, and their initial transverse momentum is negligible. The conservation of canonical momentum $p_y+a_0m_ec$ leads to synchronized oscillation of $p_y$, where $m_e$ is the electron mass, and $c$ is the speed of light.  

The right column of Fig.~\ref{fig:distr} show the longitudinal momentum distributions between $\unit[0.16]{ps}$ and $\unit[0.2]{ps}$. The pairs start with unidirectional propagation in the $+x$ direction at $t=\unit[0.16]{ps}$, illustrated in Fig.~\ref{fig:distr}(a). They begin to show negative longitudinal momenta before  $t=\unit[0.18]{ps}$ in Fig.~\ref{fig:distr}(d), indicating partial pair reflection in the region  near $x=0$. It corresponds to the cavitation of pair sphere in Figs.~\ref{fig:den}(a) and (b).
The distribution, however, does not show bump-on-tail distribution, which, together with the finite interaction time, eliminates the possibility of two-stream instability. 

\begin{figure}[th]
	\centering
	\includegraphics[width=\linewidth,valign=t]{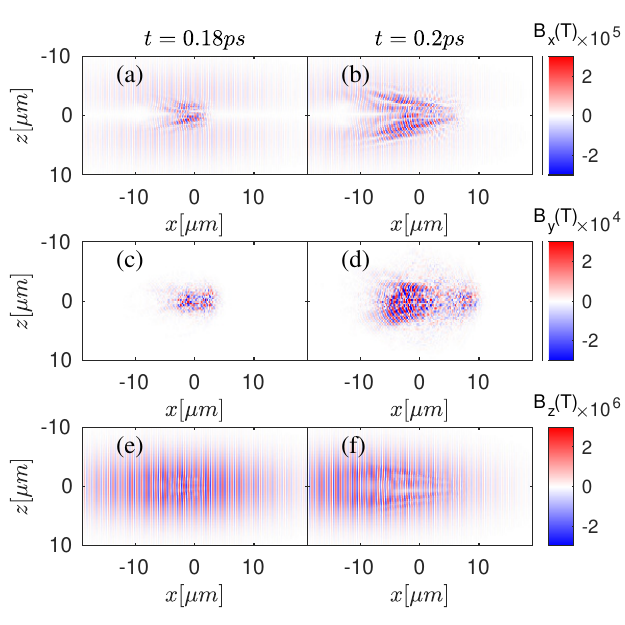}
	\caption{Magnetic field $B_x$ (a), (b), $B_y$ (c), (d), and $B_z$ (e), (f) in the $y=0$ cross section at $t=\unit[0.18]{ps}$ (left column), and $\unit[0.2]{ps}$ (right column), respectively. } 
	\label{fig:Bfield}
\end{figure}

To rule out the onset of Weibel instability, we next present the magnetic field profile in the $y=0$ cross section in Fig.~\ref{fig:Bfield}. Indeed, the transverse component $B_y$ does not show filament structures. However, filament structures are observed in the longitudinal field $B_x$. They also manifest in the decrease of laser field amplitude $B_z$ in the regions of pair filaments. Note that the background fields $B_x$ and $B_z$ extending between $z=\unit[-10]{\mu m}$ and $z=\unit[10]{\mu m}$ are the intrinsic laser field~\cite{Salamin_PRSTAB2002}, and more details can be found in the Appendix.

The decrease of the $B_z$ field in the region of pair filaments is an indication of ponderomotive filamentation~\cite{Kaw_1973}, which is supported by two main features of the pair filamentation process illustrated in Fig.~\ref{fig:den}. First, the formation of filaments is strongly associated with its interaction with the laser field rather than a counterpropagating plasma. As seen in Figs.~\ref{fig:den}(c) and (d), the forward moving plasma shell and backward moving fringes do not overlap when the fringes condensate to filaments. 

Second, the filaments exhibit higher contrast in the $z$ direction, which is perpendicular to the laser polarization direction. The anisotropic filamentation is associated with Thomson scattering of the linearly polarized laser when the pair plasma dimension is smaller than the laser wavelength. The laser drives fundamental pair oscillation in the polarized direction $\hat{y}$ and second harmonic oscillation in the longitudinal direction $\hat{x}$. But only the fundamental mode beats with the input laser and it emits most strongly in the orthogonal transverse direction $\hat{z}$. Thus, fringes in the polarization direction $\hat{y}$ are suppressed. Laser scattering in the $\hat{y}$ direction only becomes strong when the plasma volume reaches near the laser wavelength and the scattering becomes Mie scattering. 
Mie scattering has a weak dependence on polarization but still has a strong dependence on the scattering angle. The anisotropic laser scattering serves as a seed for the quickly growing pair filamentation. The laser ponderomotive force pushes the plasma to regions of lower laser intensity, evident in Figs.~\ref{fig:den}(e) and (f). The plasma density inhomogeneity  refracts the laser to regions with lower density and hence leads to filamentation instability.

\begin{figure}[b]
	\centering
	\includegraphics[width=\linewidth,valign=t]{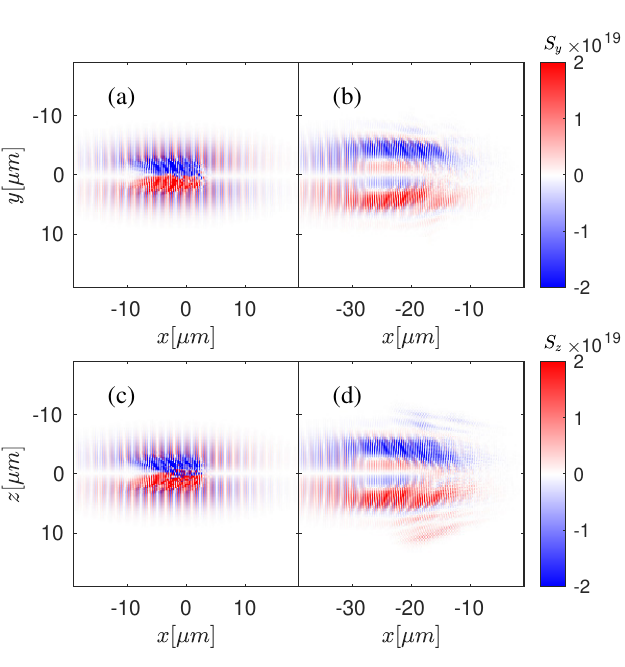}
	\caption{Poynting vector (in unit $\unit{W m^{-2}}$) $S_y$ in the $z=0$ cross section (a) , (b) and $S_z$ in the $y=0$ cross section (c), (d) at $t=\unit[0.18]{ps}$ (a), (c), and $\unit[0.26]{ps}$ (b), (d), respectively. } 
	\label{fig:dS}
\end{figure}

Mie scattering of the laser can be visualized using the Poynting vectors plotted in Fig.~\ref{fig:dS}. As soon as pair plasma is formed at $t=\unit[0.18]{ps}$, the Poynting vectors in Figs.~\ref{fig:dS}(a) and (c) show divergences in both transverse directions $\hat{y}$ and $\hat{z}$. As the filamented pairs copropagate with the laser, they continuously diffract the wave which, shown in Figs.~\ref{fig:dS}(b) and (d), deviates from the laser pulse.  Each panel of Fig.~\ref{fig:dS} also shows a low amplitude $S$ beyond the region of the filaments. They illustrate the transverse energy outflow of the Gaussian laser pulse $S_r\propto |E|^2r/R(x)$ where $R(x)$ is the radius of curvature of the beam at $x$. 
Compared to the background Poynting vector for the input laser (more details can be found in the Appendix), the diffraction modes are an order of magnitude stronger. The diffraction modes exhibit filament structure mainly in the $x$-$z$ plane, signifying the role of polarization of Thomson scattering. 


\section{Laser energy scattering} \label{sec:scattering}

\begin{figure}[b]
	\centering
	\includegraphics[width=\linewidth,valign=t]{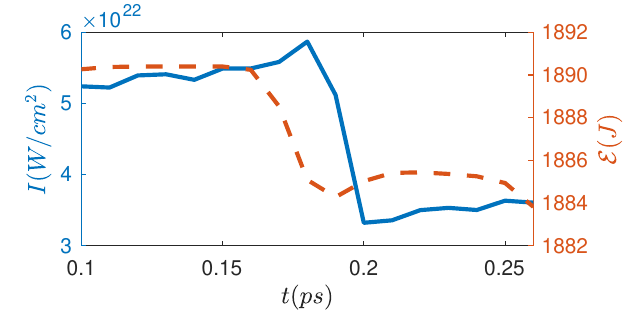}
	\caption{Evolution of the peak laser intensity (blue solid) and total energy (red dashed). } 
	\label{fig:inten}
\end{figure}

The pair filamentation has important implications to the QED cascade. It focuses the pairs in a few filaments of diameters similar to the laser wavelength, which otherwise expand continuously due to transverse laser ponderomotive force. The small filaments cause scattering of laser to large angles. Because the filaments are formed in the region of maximum laser intensity, they induce scattering of the laser energy which can be observed in the tail of the laser in Fig.~\ref{fig:sch} and in Figs.~\ref{fig:dS}(b) and (d). More clearly, the peak laser intensity, illustrated as a blue solid curve in Fig.~\ref{fig:inten}, shows a $40\%$ decrease during the pair reflection between $t=\unit[0.18]{ps}$ and $\unit[0.2]{ps}$. The figure also presents the total laser energy $\mathcal{E}=(\epsilon_0 E^2+B^2/\mu_0)/2$ as a red dashed curve, which shows only a $0.3\%$ decrease due to driving the pair oscillation and reflecting the pairs. Here $\epsilon_0$ and $\mu_0$ are the vacuum permittivity and permeability, respectively. Thus, the significant drop in peak laser intensity combined with approximately constant pulse energy comports with the theory of  laser energy scattering. 

The process of laser scattering is closely related to the laser frequency upshift when the mediating plasmas change density or Lorentz factor. In the problem of laser frequency upshift, the plasma is assumed homogeneous in the transverse direction. The transverse current thus strictly emits in the laser propagation direction or antiparallel to it. The emission couples to the original laser field to cause laser frequency upshift. However, pair filamentation breaks the symmetry and the oscillating pairs emit to the whole space due to finite current dimension.  


We can analytically estimate the rate of laser intensity decrease. The transverse current carried by the created pairs is 
\begin{align} \label{eq:J}
	\bm{J} &= \int_0^t d t' \frac{2\partial n_p}{\partial t'} \int_{t'}^t \frac{e^2}{\gamma(t'')m_e} \bm{E}(t'') d t'' \nonumber \\
	&= \frac{2e^2}{m_e}  \int_0^t \frac{n_p \bm{E}}{\gamma} dt'. 
\end{align}
We only consider the current oscillating at the laser frequency $\omega$ because the resonance causes nonreciprocating laser energy scattering. 
The pair density $n_p$ is dependent on both time and space. In the beginning of QED cascade, $n_p$ can be modeled as a $\delta$ function in space. Because the pairs oscillate synchronously, they behave as a relativistic dipole to induce Thomson scattering\footnote{The scattering is inelastic because we focus on photon emissions with the same frequency which can only beat with the input laser to create the intensity fringes.}. The angle dependence of the scattered light changes from $\sin^2\theta_1$ at $\gamma\sim0$ to $[(1-\cos\theta_2)^2 - \sin^2\theta_2\cos^2\phi/\gamma^2] /(1-\cos\theta_2)^5$ at $\gamma\gg 1$, where $\theta_1$ and $\theta_2$ are the angles between the scattered light and the laser polarization direction $\hat{y}$ and the laser propagation direction $\hat{x}$, and $\phi$ is the angle between $\hat{y}$ and the plane of the scattering direction and $\hat{x}$. When the pair volume grows to the scale of $\lambda^3$, the scattering needs to be described by the Mie theory. It is strongly dependent on the scattering angle and requires numerical treatment. But because both Thomson scattering and Mie scattering are anisotropic, the scattered light beats with the input laser to cause filamentation of the pair plasmas.

According to the Poynting theorem, the decrease of laser energy density $U$ is described with $d U/dt = {-\nabla\cdot \bm{S}} - {\bm{J}\cdot\bm{E}}$, where $U\equiv 	\epsilon_0 |E|^2$ in dispersive media. Note, however, that because Poynting flux $\bm{S}$ of the scattering arises from the emission of  the transverse current $\bm{J}$, the decrease of energy density $U$ can be found by analyzing the term ${\bm{J}\cdot\bm{E}}$. 
We describe the laser field as $\bm{E} = \bm{E}_0\cos\varphi$ and the pair density as $n_p\Theta(\varphi- \varphi_0)$ where $\varphi = \omega t-\bm{k}x$ and $\varphi_0=0$ is the instant of pair stopping at the laser peak. The transverse current can be expressed as $\bm{J} \cong 2 \epsilon_0\omega n_p\bm{E}_0\sin\varphi \Theta(\varphi- \varphi_0) /( n_c\gamma)$, where $n_c$ is the critical density of frequency $\omega$. Thus, the laser energy density decreases as 
\begin{equation}
	\frac{d U}{d\varphi} =- \frac{2\epsilon_0 n_p}{\gamma n_c} E_0^2 \sin\varphi \cos\varphi\Theta(\varphi- \varphi_0) . 
\end{equation}
The differential equation shows that energy is transferred from the laser to the transverse current when $0<\varphi<\pi/2$ and is transferred back to the wave when $\pi/2<\varphi<\pi$. But the latter process radiates to the whole space and hence, its contribution to the input laser pulse can be neglected. Therefore, the decrease of peak laser energy density near the pair stopping point can be found by taking the average of $dU/d\varphi$ in half a cycle
\begin{equation}
	\left\langle \frac{dU}{d\varphi} \right\rangle= -\frac{n_p}{\gamma n_c} \langle U\rangle .
\end{equation}
The result shows that the laser energy decreases exponentially at a rate dependent on the effective plasma density $n_p/\gamma$. 
For the presented simulation, the pair plasma has a length of $\sim12\lambda$ with $n_p/(\gamma n_c)\sim 6.7\%$. It results in a laser energy density decrease of $\sim50\%$, which agrees well with simulation results. Note that the frequency upshift process in a strict 1D scenario also causes a decrease of laser  energy density, but the decrease scales with $1/\omega$ which is an order of magnitude lower than $40\%$ decrease as obtained in the simulation. 


\section{Conclusions and discussions} \label{sec:conc}
In conclusion, we investigate the formation of pair filamentation in an electron-beam-driven QED cascade through a 3D QED PIC simulation. The pair filamentation is observed in the ``pair reflection'' regime, in which an above-threshold laser decelerates the pair plasmas and reverts their propagation direction. The simulation reveals the development of pair filaments in the longitudinal direction when the laser traverses through the pair plasma. Using a linearly polarized laser, we observe that  the pair density first exhibits periodic modulation in the plane perpendicular to the polarization direction. Within a few laser cycles, the density modulation condenses into filaments with a diameter similar to the laser wavelength. Importantly, this timescale extends beyond that of the streaming instabilities. Furthermore, considering the absence of a robust transverse magnetic field perpendicular to the laser, we dismiss the possibility of two-stream instability or Weibel instability in this scenario.

We attribute the origin of pair filamentation to the laser ponderomotive force. The initially small pair plasma induces highly anisotropic scattering of the laser, creating interference patterns with the incident laser and establishing a pressure gradient within the pair plasma. Pairs are expelled from regions of high laser intensity, and pair cavitation gives rise to locally elevated refractive indices that focus the laser pulse. The inhomogeneous plasma density also contributes to the refraction of laser energy. Consequently, filamentation takes shape in the region where dense pair plasma is stopped.
As the reflected pairs and the laser pulse copropagate, their slippage is diminished, facilitating the rapid growth of pair filaments. This reduced slippage enables efficient coupling between the reflected pairs and the laser pulse, augmenting the development of pair filaments in this dynamic interplay.

Laser polarization critically influences the formation of pair plasma filamentation. In the initial stage of interaction, the small plasma oscillates synchronously to cause stronger scattering in the direction perpendicular to the polarization direction. Because the scattering is negligible in the polarization direction, a linearly polarized laser only creates 2D gratings in the initial stage when the pair plasma is smaller than the laser wavelength. 
As the plasma volume grows, the scattering becomes less polarization dependent.

The rapidly expanding plasma volume in the beam-driven QED cascade sets itself apart from ponderomotive filamentation observed when a low-intensity laser traverses stationary plasmas. In the QED cascade, the small pair plasma does not only directly seed the filamentation, but also accentuates the instability through causing Mie scattering. The beat of the scattered light and input laser leads to strong modulation of pair density. A full analytical model that describes the coupling requires carefully treating both ultrarelativistic particle motion and also complicated Mie scattering in multidimensions. Thus, the filamentation growth rate of the pair plasma significantly deviates from known expressions of the ponderomotive filamentation instability~\cite{Kruer1985}.

The pair filamentation has important implications to the joint problem of creating and observing QED plasmas. First, filamentation focuses pairs and creates high pair density regions inside the laser field. Compared to homogeneously distributed pairs, filamentation reduces the total pair number needed to reach high pair density. Second, despite being highly localized, the dense pair plasma interacts with the most intense part of the laser. The strong scattering of laser energy due to small filament dimensions causes a dramatic decrease in laser intensity within the filament length. All the scattered laser energy is redistributed and can  be detected at large angles. 
Because the scattering arises from the interaction of collective pair motion and the laser fields, it serves as one more signature of QED plasma effect.

\begin{acknowledgments}
	This work was supported by NSF Grant No.  PHY-2206691. 
\end{acknowledgments}

\setcounter{figure}{0}                       
\renewcommand\thefigure{\arabic{figure}*}    
\setcounter{equation}{0}                       
\renewcommand\theequation{\arabic{equation}*}    

\section*{Appendix}

The linearly polarized Gaussian laser pulse contains fields in multiple directions due to its finite waist size, $w_0=\unit[5]{\mu m}$. With its electric field polarized in the $y$ direction, the electric and magnetic fields in vacuum can be written to the second order of $(w_0/x_R)\approx0.05$ as~\cite{Salamin_PRSTAB2002}
\begin{align}
	E_x &= \tilde{E} \frac{y}{x_R} \frac{w_0}{w(x)} \cos(\varphi + 2\varphi_G - \varphi_R).  \\
	E_y &= \tilde{E} \Big\{ \sin(\varphi + \varphi_G)  + \frac{w_0^2}{x_R^2} \Big[ \frac{y^2}{w^2(x)} \sin(\varphi + 3\varphi_G) \nonumber \\
	&\qquad\qquad  - \frac{(y^2+z^2)^2}{4w_0 w^3(x)} \sin(\varphi + 4\varphi_G) \Big]\Big\}, \\
	E_z &= \tilde{E} \frac{yz}{x_R^2} \frac{w_0^2}{w^2(x)} \sin(\varphi + 3\varphi_G), \\
	B_x &= \frac{\tilde{E}}{c} \frac{z}{x_R} \frac{w_0}{w(x)} \cos(\varphi + 2\varphi_G),  \\
	B_y &= 0, \\
	B_z &= \frac{\tilde{E}}{c}  \Big\{ \sin(\varphi + \varphi_G)+  \frac{w_0^2}{x_R^2} \Big[ \frac{y^2+z^2}{2w^2(x)} \sin(\varphi + 3\varphi_G) \nonumber \\
	&\qquad\qquad   - \frac{(y^2+z^2)^2}{4w_0 w^3(x)} \sin(\varphi + 4\varphi_G) \Big]\Big\}, 
\end{align}
where $k=-2\pi/\lambda$, $x_R = kw_0^2/2$ is the Rayleigh range, $w(x) = w_0\sqrt{1+(x/x_R)^2}$ is the spot size at distance $x$, $\varphi = \omega t - kx - k(y^2+z^2)/(2R)$, $R=x+x_R^2/x$, $\varphi_G = -\arctan(x/x_R)$ is the Gouy phase, and $\tilde{E} = E_0 [w_0/w(x)]\exp[ -(y^2+z^2)/w^2(x) ] \exp[-(t-x/c)^2/\tau^2]$ is the envelope of the Gaussian beam, with $\tau=\unit[50]{fs}$.

\begin{figure}[thb]
	\centering
	\includegraphics[width=\linewidth,valign=t]{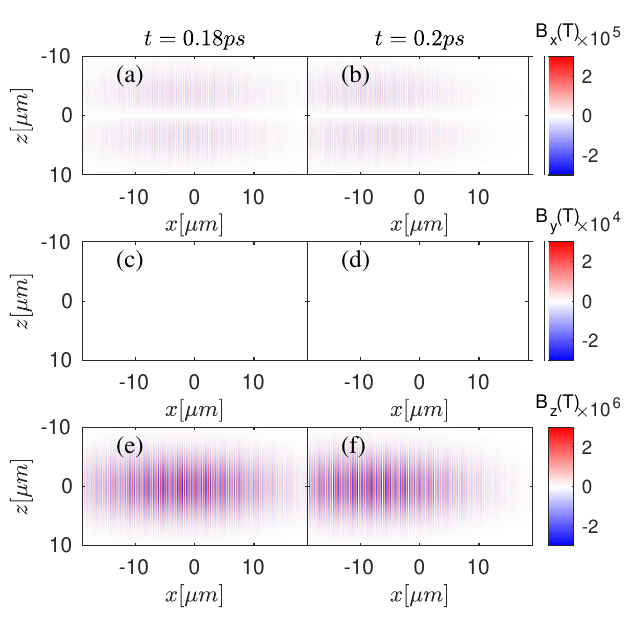}
	\caption{The laser magnetic field $B_x$ (a), (b), $B_y$ (c), (d), and $B_z$ (e), (f) in the $y=0$ cross section without interacting with electron beam at $t=\unit[0.18]{ps}$ (left column), and $\unit[0.2]{ps}$ (right column), respectively. } 
	\label{fig:Bfield_ref}
\end{figure}

\begin{figure}[thb]
	\centering
	\includegraphics[width=\linewidth,valign=t]{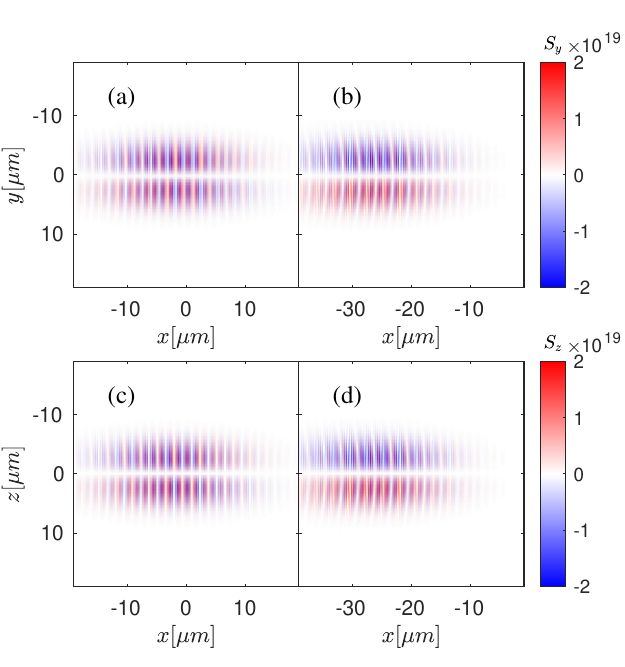}
	\caption{Poynting vector (in unit $\unit{W m^{-2}}$) $S_y$ in the $z=0$ cross section (a), (b) and $S_z$ in the $y=0$ cross section (c), (d) of the laser without interacting with electron beam at $t=\unit[0.18]{ps}$ (a), (c), and $\unit[0.26]{ps}$ (b), (d), respectively. } 
	\label{fig:dS_ref}
\end{figure} 

\begin{figure}[thb]
	\centering
	\includegraphics[width=\linewidth,valign=t]{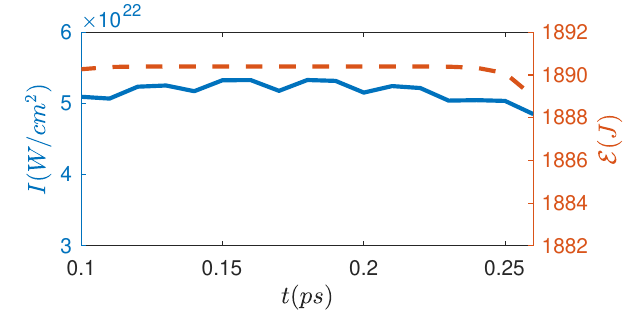}
	\caption{Evolution of the peak laser intensity (blue solid) and total energy (red dashed). } 
	\label{fig:inten_ref}
\end{figure}

Figure~\ref{fig:Bfield_ref} shows the magnetic fields of laser in vacuum for direct comparison with Fig.~\ref{fig:Bfield}.

Figure~\ref{fig:dS_ref} shows the Poynting vectors of laser in vacuum for direct comparison with Fig.~\ref{fig:dS}. The decrease of energy at $t=\unit[0.26]{ps}$ is caused by laser exiting the simulation window. 

Figure~\ref{fig:inten_ref} shows the evolution of the peak laser intensity and total energy in vacuum for direct comparison with Fig.~\ref{fig:inten}.

\bibliography{Upshift}

\end{document}